\documentclass{article}

\input pz.sty
\input epsf.sty
\usepackage{longtable}

\usepackage{amssymb,amsmath,epsfig}

\begin{document}

\begin{center}

\PZtitletl{Variability of Hot Supergiant IRAS~19336--0400} {in the
Early Phase of its Planetary Nebula Ionization}

\PZauth{V.P.Arkhipova, M.A.Burlak, V.F.Esipov,} {N.P.Ikonnikova*,
G.V.Komissarova}

\PZinsto{Moscow State University, Sternberg State Astronomical
Institute, Moscow}

\begin{abstract}

{\sloppy We present photoelectric and spectral observations of a
hot candidate protoplanetary nebula -- early B-type supergiant
with emission lines in spectrum -- IRAS~19336--0400. The light and
color curves display fast irregular brightness variations with
maximum amplitudes $\Delta V=0^{m}.30$, $\Delta B=0^{m}.35$,
$\Delta U=0^{m}.40$ and color-brightness correlations. By the
variability characteristics IRAS~19336-0400 appears similar to
other hot protoplanetary nebulae. Based on low-resolution spectra
in the range $\lambda$ 4000-7500 \AA\ we have derived absolute
intensities of the emission lines H$\alpha$, H$\beta$, H$\gamma$,
[SII], [NII], physical conditions in gaseous nebula:
$n_e=10^4$~cm$^{-3}$, $T_{e}=7000\pm1000$~K. The emission line
H$\alpha$, H$\beta$ equivalent widths are found to be considerably
variable and related to light changes. By \textit{UBV}--photometry
and spectroscopy the color excess has been estimated:
$E_{B-V}$=0.50--0.54. Joint photometric and spectral data analysis
allows us to assume that the star variability is caused by stellar
wind variations.

}

\bigskip

{\sloppy {\it {Key-words}}: proto-planetary nebulae, photometric
and spectral observations, light variability

}

$^*$ Send offprint requests to: ikonnikova@gmail.com
\end{abstract}

\end{center}

\section*{Introduction}

According to our current knowledge, after the superwind has
ceased, a star with a main sequence initial mass between 1 and
8~$M_{\odot}$ departs from the tip of the asymptotic giant branch
(AGB) leaving a carbon-oxygen core with a mass of
0.5--0.8~$M_{\odot}$ surrounded by extended gaseous nebula
imitating the supergiant characteristics (luminosity,
gravitational acceleration). The object is also surrounded by a
dust shell becoming optically thinner as expanding. Theoretical
calculations predict the star to evolve through the
post-asymptotic (post--AGB) phase, with its bolometric luminosity
being constant while its radius shrinking and effective
temperature rising. When the temperature is high enough
($T_{\textrm{eff}} \approx 20\,000$~Ê) to ionize the circumstellar
nebula, there appear emission lines, and the object becomes
detectable by spectral sky surveys owing to the presence, first of
all, of H$\alpha$ emission line. Hot post--AGB stars, or
proto-planetary nebulae (PPNe), are the immediate progenitors of
the central stars of planetary nebulae (CSPN).

{\sloppy Hot candidate PPNe possess dust shells with temperatures
100\,K\,$\!<\!T_{\textrm{dust}}\!<$\,250\,K, display spectra of
early B--type with signs of a supergiant and emission lines, and
are usually located outside the Galactic plane.

}

The object IRAS~19336--0400 ($\alpha=19^{\textrm{h}}
36^{\textrm{m}}17^{\textrm{s}}.5; \delta= -03^{\circ} 53' 25''.3$
(2000)) totally satisfies all these requirements. The source is
located at high galactic latitude ($b=-11^{\circ}$.75) far from
star formation regions.

The star was first identified in the survey of Stephenson and
Sanduleak (1977) as an object with $V = 12^{m}.5$, OB--type
spectrum, H$\beta$ emission and was designated as SS~441. Later
Downes and Keyes (1988) examined its optical spectrum in the range
$\lambda$~3700--7500~\AA\ in more detail and pointed out the
presence of strong Balmer lines, low excitation forbidden lines of
[OII], [NII], [SII] and the absence of HeI, HeII and [OIII] lines.
The authors estimated the general appearance of the spectrum to be
consistent with the class BQ[~].

IRAS mission measured the flux from SS~441 in far-infrared region
(12 -- 100~$\mu$m) emitted by cold dust with $T_{\textrm{dust}}$ =
175\,K (Preite-Martinez, 1988). In IRAS Catalogue the object was
designated as IRAS 19336--0400 and based on its location in the IR
color-color diagram [12]--[25], [25]--[60] Preite-Martinez put it
into the catalogue of possible planetary nebulae.

The object was detected in radio emission at 6~cm (Van de Steene
and Pottasch 1995). After Van de Steene et al. (1996) had
described its optical spectrum and analyzed IR and radio
observations the object was included by Kohoutek (2001) into the
catalogue of planetary nebulae under the name of PN~034--11.1.

Analyzing IRAS~19336--0400 spectrum in the range 5 -- 40~$\mu$m
obtained by Spitzer Space Telescope (Cerrigone et al. 2009)
allowed to conclude that the object's dust shell had mixed
carbon--oxygen chemistry, since the spectrum displayed both a
broad 10~$\mu$m feature attributed to the SiO molecule and 6.2 and
7.7~$\mu$m details due to polycyclic aromatic hydrocarbons (PAHs).
A possible reason for the mixed chemistry may be binarity that
favors the formation of circumbinary disk with crystalline
silicates after the star has been enriched with carbon on AGB
(Bregman et al. 1993, Waters et al. 1998).

All the data available up to the beginning of our research
indicated that IRAS~19336--0400 belongs to the objects in
post--AGB evolutionary stage. Hot PPNe in the early phase of
envelope ionization are known to display activity in the form of
photometric and spectral variability (Arkhipova et al. 2006). We
hoped to detect such variability in the IRAS~19336--0400 behavior,
and for this purpose we carried out photometric and spectral
observations for several years. Hitherto there was no published
information concerning photometric behavior of the star, and there
existed discrepant data on emission line absolute intensities.

In 2006 IRAS~19336--0400 was included into the hot PPNe
variability search and study programme being executed in SAI for
more than 20 years. As a result of $UBV$--observations there has
been reliably detected fast light variability, and one chapter of
the current paper is devoted to its analysis.

Spectral investigations of the star based on the observations that
we carried out with 125--cm telescope at SAI Crimean Laboratory in
2008--2011 allowed us to measure emission line absolute fluxes, to
derive physical conditions in the nebula, to detect emission line
equivalent widths variations. And also we compared our spectral
observations with the data of other researchers.

\section*{$UBV$--observations of IRAS~19336--0400}

In 2006 we started systematic photometric observations of
IRAS~19336--0400. Measurements were carried out on the 60--cm
telescope Zeiss--1 at SAI Crimean Station with the use of
photoelectric $UBV$--photometer constructed by V.M.~Lyuty (1971).
Photometric system of the photometer is close to that of Johnson.
$UBV$--magnitudes of the comparison star HD~184790 ($V=8.^{m}125$,
$B=8.{^m}293$, $U=7.{^m}966$) were taken from Crawford et al.
(1971). For IRAS~19336--0400 the precision of IRAS~19336--0400
observations of $\sigma\sim 0.^{m}02$ was obtained for each of
three filters.

In six years of observations we got 155 brightness estimates for
IRAS~19336--0400. Table~1 lists the $UBV$--magnitudes of
IRAS~19336--0400 for the period 2006--2011, Fig.~1 shows light and
color curves of the star. The star displays fast irregular
brightness and color variations with maximum amplitudes $\Delta
V=0.{^m}30$, $\Delta B=0.{^m}35$, $\Delta U=0.{^m}40$, $\Delta
(B-V)=0.^{m}15$, $\Delta (U-B)=0.^{m}25$. Light variability may be
considered real since $3\sigma$--threshold is exceeded several
times. According to our measurements the average brightness values
of IRAS~19336--0400 in 2006--2011 turn out to be
$<\!V\!>=13.^{m}15$, $<\!B\!>=13.^{m}47$, $<\!U\!>=12.^{m}72$. In
2011 (JD~2455715--2455795) the light variation amplitude did not
exceed $0.^{m}2$ in all bands.

The character of IRAS~19336--0400 light variations can be traced
by observations obtained during three moonless seasons in 2008,
when the star was measured almost every night. Fig.~2 shows the
light curves of the star in the time interval JD 2454646-2454717.
According to these observations the star brightness changes
regularly from night to night with characteristic time from 7 to
13 days in different seasons. We failed to determine period both
in the total data set and in the subsets of single seasons.

{\sloppy To explain the photometric instability of
IRAS~19336--0400 we consider two hypotheses, variable stellar wind
and/or pulsations, assumed by Handler (2003) to be the most likely
reasons for the group of "cold"\ central stars of young planetary
nebulae
--- ZZ~Lep--type variables, whose photometric variability is
similar to that of some hot post--AGB stars.

}

We have analyzed color--magnitude diagrams including the total set
of IRAS 19336-0400 observations (Fig.~3). Colors correlate with
brightness, $B-V$ increasing and $U-B$ decreasing when the star is
getting brighter. Such a color behavior can not be explained by
the temperature variations due to pulsations when brightening is
followed by both colors, $U-B$ and $B-V$, becoming bluer. ZZ~Lep
is also observed to be redder in $B-V$--color when brighter
(Handler 2003).

On the other hand Boyarchuk and Pronik (1965) pointed out for Be
stars (star+gaseous envelope system) that the envelope radiation
always increases $B-V$ and decreases $U-B$ simultaneously.

It's natural to suppose that, if the star brightens because of the
stellar wind enhances, and thus the envelope radiation increases
its contribution primarily to the hydrogen continuum, then $B-V$
should increase and $U-B$ should decrease.

Accordingly to its spectral type B1I (Vijapurkar et al. 1998) we
have assumed for IRAS~19336--0400 normal color $(B-V)_{0}=-0.19$
in the calibration of Straizys (1982) and defined the extinction
value $E_{B-V}=0.50\pm0.07$. It's worth mentioning that correcting
the mean color $U-B=-0.75$ with the obtained value of $E_{B-V}$
leads to $U-B=-1.15$ that is 0.15 smaller than $(U-B)_{0}=-1.0$
for a B1--supergiant. So there exists an appreciable excess of
radiation in $U$--band that argues in favor of the assumption of
additional Balmer continuum emitted by a gaseous envelope.

We have examined separately two--color $(U-B),(B-V)$ diagram for
the observations of 2008, the light curves for which are presented
in Fig.~2. In Fig.~4 we plot colors of the star both observed and
corrected for interstellar reddening with $E_{B-V}=0.50$. There
are also shown the sequence of supergiants and theoretical
$(U-B)-(B-V)$ relations for optically thin in Balmer continuum
hydrogen plasma with $n_{e}=10^{10}$~cm$^{-3}$ and
$T_{e}=10\,000$~K and for plasma with $T_{e}=15\,000$~K and
variable optical thickness in Balmer continuum taken from Chalenko
(1999). One can see that almost all point corrected for reddening
lie above the sequence of supergiants and the star oscillation on
the diagram upwards to the right and downwards to the left is
broadly in line with the direction of hydrogen continuum opacity
change probably associated with stellar wind density variability.

We consider unsteady stellar wind to be the main cause of
photometric variability. Its origin is to be associated with
density inhomogeneities in the outer levels of the star, a low
mass supergiant passed through the stage of hydrogen and helium
burning on AGB.

{\sloppy Photometric variability of IRAS~19336--0400 appears very
similar to that of other hot post--AGB stars. As we reported
previously (Arkhipova et al. 2006), B--supergiants with
IR--excess: V886~Her, LS\,IV $-12^{\circ}111$, V1853~Cyg,
IRAS~19200+3457 and IRAS~07171+1823 also display fast irregular
photometric variability with amplitudes of $0.^{m}2-0.^{m}4$ in
$V$--band and color--brightness correlations analogous to
IRAS~19336--0400.

For four of the six hot PPNs mentioned above (except for
IRAS~19200+3457 and IRAS~19336--0400) there exist spectra of high
resolution that allowed to detect P~Cyg profiles of HeI and HI
recombination lines indicating mass loss in the stars. In the
cases when spectral monitoring was carried out (i.e. for
V1853~Cyg) significant variability of HeI $\lambda$6678~\AA\ line
profile and variability of atmospheric absorption line radial
velocities was found (Garc\'\i a-Lario et al. 1997).

}

\section*{Spectral observations of IRAS~19336--0400}

Earlier the optical spectrum of IRAS~19336--0400 was investigated
several times. Carrying out spectral observations for objects with
H$\alpha$ emission Downes and Keyes (1988) obtained for
IRAS~19336--0400 a low resolution ($\sim$13\AA) spectrum where one
could see a featureless slightly red sloping continuum typical for
a B--class star and forbidden emission lines of [OII]
$\lambda3727$, [NII] $\lambda6584$, [SII] $\lambda6717, 6731$,
found in low excitation PNe. Later performing the project to
identify new PNe Van de Steene et al. (1996) measured emission
line fluxes and estimated the value of interstellar extinction
($A_{V}=1.7$). Higher resolution ($\sim$8\AA) allowed the authors
to identify weak absorption lines of CIII $\lambda4648$, Na~I,
He~I. The object was confirmed to be a PN with a B--type central
star. Vijapurkar et al. (1998), Parthasarathy et al. (2000)
specified the central star spectral type (B1Iape) and confirmed
low excitation of the nebula. Having measured emission line
intensities Pereira and Miranda (2007) derived interstellar
extinction, electron temperature and density in the nebula, some
relative abundances (N,O,S) and based on the results suspected
IRAS~19336--0400 to belong to type III PNe according to the
classification of Peimbert (1990).

We observed IRAS~19336--0400 spectroscopically in 2008--2011 at
the 125--cm reflector of SAI Crimean Station. We used a
diffraction spectrograph with 600~lines/mm grating. The slit width
was 4$''$.

The detector was a ST--402 CCD ($765\times510$ pixels of
$9\times9\mu$m). Spectral resolution (FWHM) was 7.4~\AA\, in the
spectral range of 4000-7200\AA.

{\sloppy The spectra were reduced using the standard CCDOPS
program and also the program SPE created by S.G. Sergeev at
Crimean Astrophysical Observatory. To calibrate fluxes the spectra
of standard stars were observed: 18~Vul, 4~Aql, $\kappa$~Aql.
Absolute spectral energy distributions for standard stars were
taken from electronic versions of spectrophotometric catalogues
(Glushneva et al. 1998, Kharitonov et al. 1988, Alekseeva et al.
1997). It's worth mentioning that the discrepancy of data for the
same spectrophotometric standard stars between different
catalogues is about 10\%\, and it is especially significant in the
blue range. While calibrating spectra we did not take into account
the air mass difference between the program star and the standard
star though sometimes it was equal to 0.4. Besides on some nights
the weather was instable. Therefore after the primary calibration
by standard star spectra we then adjusted derived spectral energy
distributions to photometric data obtained on the same nights.

}

Fig.~5 shows the spectrum of IRAS~19336--0400 obtained on July 4,
2008. One can easily see emission lines of hydrogen, [NII], [SII],
[OI], generated in the nebula and superposed on the slightly red
sloping continuum of the central star with faint absorption lines
of HeI $\lambda4921$, $\lambda5876$, $\lambda6678$ and
$\lambda7065$, the absorption lines of NaI $\lambda5892$ and CIII
$\lambda4648$ are also present.

{\sloppy We have measured absolute fluxes of nebula emission lines
by integration over line profiles. Measuring uncertainties come
from natural noise, instrumental errors, calibration errors,
uncertainty of the underlying continuum position. We estimate the
intensity error to be about 10\% for stronger lines and about 30\%
for [NII] $\lambda5755$. Our accuracy and spectral resolution
appeared insufficient to investigate absorption lines behavior.

Table~2 lists observed relative emission line intensities scaled
to $F_{\textrm{H}\beta}=100$, absolute intensity of H$\beta$ line
and the [SII] line intensities ratio
$R(\textrm{SII})=\frac{\lambda6717}{\lambda6731}$, obtained in
this work and averaged over the observation period and the
quantities found by Van de Steene et al. (1996) and by Pereira,
Miranda (2007). Van de Steene et al. (1996) did not give
explicitly the measured absolute H$\beta$ intensity but in the
paper there is a graphical spectrum presentation in absolute
energy units that allowed us to assess the required quantity for
comparison. Our intensities both relative and absolute are in
reasonable agreement with those of Van de Steene et al. (1996).
The same is true for relative intensities given by Pereira,
Miranda (2007) but not for the absolute ones: both H$\beta$ line
intensity and the continuum level obtained in our study are 4--5
times higher than that from Pereira, Miranda (2007). If we managed
to carry out both photometric and spectral observations
simultaneously, we adjusted the calibration of our spectra to $B$
and $V$ magnitudes of IRAS~19336--0400 obtained on the same
nights. The absolute fluxes of Pereira, Miranda (2007) seem too
low even if to take into consideration light variability of
IRAS~19336--0400 and calibration errors.

}

From the intensity ratio for the first two Balmer lines we
estimated the interstellar extinction for IRAS~19336--0400
assuming the dereddened decrement from Hummer, Storey (1987)
($\textrm{H}\alpha/\textrm{H}\beta=2.9$ for $n_e=10^4$~cm$^{-3}$,
$T_e=7.5\times10^3$~K) and the extinction law of Seaton (1979).
The value of $c_{\textrm{H}\beta}$, characterizing the
interstellar extinction in $\textrm{H}\beta$ line and referred to
color excess as $c_{\textrm{H}\beta}=1.47E_{B-V}$, appeared to be
0.8 and in good agreement with the results of Van de Steene et al.
(1996) and also of Pereira, Miranda (2007) with some remarks
(Table~2).

{\sloppy From the intensity ratios of forbidden lines
$R(SII)=\lambda6717/\lambda6731$ and
$R(NII)=(\lambda6548+\lambda6584)/\lambda5755$ we have derived the
values of electronic temperature and density in the nebula
(Table~2) that are close to the quantities of Pereira, Miranda
(2007). $R(SII)$ value is close to the limit one, attained at high
density, therefore our $N_e$ estimate is valid only up to an order
of magnitude.

Firstly we planned to study the spectral variability of
IRAS~19336--0400 and its relation to the photometric variability.
The emission component of IRAS~19336--0400 optical spectrum
consists of several nebula lines weak enough, relative to the
stellar continuum, to affect the summary brightness in
$UBV$--bands. So one may assume that calibration by photometry
does not prevent from detecting emission line variability. But
during four years of observations we did not detect emission line
intensity variations beyond the bounds of error. It may be
explained by the fact that fast irregular variability of the
central star barely affects the emission line spectrum generated
in the gaseous nebula. Thereby we consider the summary spectrum of
IRAS~19336--0400 to consist of constant emission component,
related to the young PN, and variable continuum of the central
star.

}

To detect continuum variability and to exclude errors arising from
calibration we studied how the equivalent widths of H$\alpha$ and
H$\beta$ emission lines vary in relevance to the star $B$ and $V$
brightness and we found inverse correlation between the equivalent
widths of given lines and brightness: the widths appear to be
smaller when the star is brighter. Fig.~6 shows the results of
simultaneous spectral and photometric observations (10 nights).

{\sloppy The detected correlation of brightness and H$\alpha$ and
H$\beta$ equivalent widths may be explained in the following way:
the star brightens because mass outflow rate increases and the
contribution of stellar wind radiation to the total continuum
luminosity becomes more prominent. The emission line equivalent
widths are the largest when the wind continuum contribution to the
total continuous spectrum is rather small (the star is faint). If
the mass outflow rate increases (the star is bright) then the
continuum emission of the wind raises the total continuum level
and the emission line equivalent widths decrease.

}

\section*{Conclusions}

The photometric and spectral monitoring of the hot PPN
IRAS~19336--0400 allowed to detect light and spectral variability
of the star.

Fast night to night irregular light oscillations with maximum
amplitudes of $\Delta V=0.^{m}3$, $\Delta B=0.^{m}35$, $\Delta
U=0.^{m}4$ have been found. In addition the connection between
light and colors has been detected: the star brightening is
followed by $U-B$ decreasing and $B-V$ increasing.

We suppose that photometric variability of the object is caused by
the sporadical mass outflow rate change ($\dot{M}$) appearing as a
variable continuum radiation excessive if compared to the spectral
energy distribution of a B1--supergiant star. This additional
continuum may be fitted by emission of hydrogen gas with electron
density $\sim\!10^{10}$ cm$^{-3}$ and temperature from 10\,000 to
15\,000~K.

Color variations also argue in favor of gaseous envelope
contribution changes and against temperature variations of the
stellar photosphere. To define reliable characteristic times of
variability it is necessary to perform photometric observations
with better time resolution than ours.

{\sloppy By low-resolution spectral observations of
IRAS~19336--0400 Balmer emission line equivalent widths have been
found to be variable with time that points out the continuum
variability while nebular emission line fluxes remain constant
within measuring accuracy. The correlation of star brightness with
nebular emission line equivalent widths has been detected and the
largest widths have been observed at the star maxima. The
correlation arises from emission line equivalent widths becoming
larger when continuum level is dropping while the star is fading.

IRAS~19336--0400 has become the sixth object in the group of
variable B--supergiants with IR--excess along with V886~Her,
LS\,IV--12$^{\circ}$111, V1853~Cyg, IRAS~19200+3457 and
IRAS~07171+1823. Astonishing similarity in photometric behavior of
these stars evidences that they have the common nature of
variability.

Treating the variability of IRAS~19336--0400 and other hot PPN
candidates we rely on the assumption that during post--AGB
evolution, when the star temperature is rising, there starts a new
phase of mass loss driven by light pressure in resonance lines
(Kwok 1993). The stellar wind of these stars is variable that is
evidenced by high resolution spectral data obtained by various
researchers for the majority of the sample stars. Spectroscopic
observations of the same quality for IRAS~19336--0400 should be
invaluable, and it is especially important to study line profiles,
to define radial velocities and their relation to the star
brightness.

}

\section*{References}
{\sloppy

\begin{enumerate}

    \item G.A. Alekseeva, A.A. Arkharov, V.D. Galkin, et al., \textit{VizieR Online Data Catalog}
    III/201 (1997).
    \item V.P. Arkhipova, N.P. Ikonnikova,
    G.V. Komissarova, and R.I. Noskova, \textit{Planetary Nebulae
    in our Galaxy and Beyond}, IAU Symp. 234.(Ed. Ian F. Corbett, Cambridge: Cambridge
Univ. Press, 2006), p. 357--358.
    \item A.A. Boyarchuk, I.I. Pronik, Izv. Krym. Astrofiz. Obs. {\bf 33}, 195
    (1965).
    \item J.D. Bregman, D. Jesse, D. Rank, et al., Astrophys. J. {\bf 411}, 794 (1993).
    \item L. Cerrigone, J.L. Hora, G. Umana, ànd C. Trigilio,
    Astrophys. J., {\bf 703}, 585 (2009).
    \item N.N. Chalenko, Astron. Zh. {\bf 76}, 529 (1999)[Astron. Rep. {\bf 43}, 459 (1999)].
    \item D.L. Crawford, J.C. Golson, and A.U. Landolt,
    Publ. Astron. Soc. Pac. {\bf 83}, 652 (1971).
    \item R.A. Downes and C.D. Keyes, Astron. J. {\bf 96}, 777 (1988).
    \item P. Garc\'{i}a-Lario, M. Parthasarathy, D. de
    Martino, et al., Astron. Astrophys. {\bf 326}, 1103 (1997).
    \item I.N. Glushneva, V.T. Doroshenko, T.S. Fetisova, et al., \textit{VizieR Online Data Catalog} III/208 (1998).
    \item G. Handler, \textit{Interplay of Periodic, Cyclic and Stochastic
    Variability in Selected Areas of the H-R Diagram} (Ed. C. Sterken, ASP Conf. Ser. 292, San Francisco:
    Astron. Soc. Pacific, 183, 2003).
    \item D.G. Hummer and P.J. Storey, MNRAS {\bf 224}, 801 (1987).
    \item A.V. Kharitonov, V.M. Tereshchenko, L.N.
    Knyazeva, \textit{VizieR Online Data Catalog} III/202 (1988).
    \item L. Kohoutek, Astron. Astrophys. {\bf 378}, 843 (2001).
    \item S. Kwok, Annu. Rev. Astron. Astrophys. {\bf 31},
    63 (1993).
    \item V.M. Lyuty, Soobshch. GAISh {\bf 172}, 30 (1971).
    \item D. Monet, A. Bird, B. Canzian, et al., \textit{USNO-A V2.0, A Catalog of Astrometric Standards.
    U.S. Naval Observatory Flagstaff Station (USNOFS) and
     Universities Space Research Association (USRA) stationed at USNOFS}(1998).
    \item M. Parthasarathy, J. Vijapurkar, and J.S.
    Drilling, Astron. Astrophys. Suppl. Ser. {\bf 145}, 269
    (2000).
    \item M. Peimbert, Rep. Prog. Phys. {\bf 53}, 1559 (1990).
    \item C.B. Pereira and L.F. Miranda, Astron. Astrophys. {\bf 467}, 1249
    (2007).
    \item A. Preite-Martinez, Astron. Astrophys. Suppl. Ser. {\bf 76}, 317 (1988).
    \item M.J. Seaton, MNRAS {\bf 187}, 73P (1979).
    \item C.B. Stephenson and N. Sanduleak, Astrophys. J. Suppl. Ser. {\bf 33}, 459 (1977).
    \item V. Straizys, {\it Metal-deficient stars} (Mokslas, Vil'nyus,
    1982).
    \item G.C. Van de Steene and S.R. Pottasch, Astron. Astrophys.,
    {\bf 299}, 238 (1995).
    \item G.C. Van de Steene, G.H. Jacoby, and S.R. Pottasch, Astron. Astrophys. Suppl. Ser. {\bf 118}, 243
    (1996).
    \item J. Vijapurkar, M. Parthasarathy, and J.S. Drilling, Bull. Astron. Soc. India {\bf 26}, 497 (1998).
    \item L.B.F.M. Waters, J. Cami, T. de Jong, et al.,
    Nature {\bf 391}, 868 (1998).

\end{enumerate}

}
\newpage
{\small
\begin{center}
\begin{longtable}{|c|c|c|c||c|c|c|c|}
\caption{$UBV$--observations of IRAS~19336--0400 in 2006--2011.}\\
\hline
JD& \textit{V} & \textit{B--V} & \textit{U--B} &JD& \textit{V} & \textit{B--V} & \textit{U--B}\\
\hline
\endfirsthead

\multicolumn{4}{c}{\tablename\ \thetable{}}\\
\hline
JD& \textit{V} & \textit{B--V} & \textit{U--B} &JD& \textit{V} & \textit{B--V} & \textit{U--B} \\
\hline
\endhead

\hline
\endfoot

\hline
\endlastfoot

\hline
2453911&   13.18&    0.27& -0.72&2454592&   13.12&    0.31& -0.75\\
2453912&   13.15&    0.34& -0.72&2454594&   13.16&    0.35& -0.79\\
2453915&   13.24&    0.26& -0.64&2454602&   13.24&    0.27& -0.71\\
2453941&   13.04&    0.32& -0.69&2454616&   13.07&    0.33& -0.67\\
2453943&   13.04&    0.35& -0.69&2454617&   13.17&    0.32& -0.78\\
2453946&   13.04&    0.34& -0.74&2454620&   13.25&    0.27& -0.74\\
2453947&   13.27&    0.30& -0.72&2454622&   13.21&    0.32& -0.65\\
2453950&   13.09&    0.36& -0.75&2454624&   13.12&    0.35& -0.81\\
2453966&   13.18&    0.28& -0.73&2454628&   13.35&    0.32& -0.69\\
2453968&   13.14&    0.36& -0.75&2454630&   13.16&    0.29& -0.76\\
2453969&   13.22&    0.33& -0.81&2454646&   13.15&    0.33& -0.79\\
2453971&   13.07&    0.31& -0.73&2454647&   13.20&    0.29& -0.77\\
2453972&   13.11&    0.31& -0.74&2454649&   13.16&    0.28& -0.71\\
2453974&   13.20&    0.30& -0.81&2454655&   13.06&    0.30& -0.67\\
2453990&   13.18&    0.30& -0.76&2454659&   13.04&    0.38& -0.77\\
2454029&   13.21&    0.30& -0.69&2454660&   13.16&    0.30& -0.75\\
2454036&   13.22&    0.28& -0.78&2454677&   13.23&    0.26& -0.66\\
2454214&   13.22&    0.30& -0.77&2454678&   13.12&    0.34& -0.77\\
2454218&   13.26&    0.27& -0.81&2454679&   13.15&    0.31& -0.72\\
2454232&   13.12&    0.32& -0.73&2454681&   13.32&    0.29& -0.73\\
2454233&   13.08&    0.36& -0.75&2454682&   13.19&    0.30& -0.81\\
2454236&   13.15&    0.34& -0.75&2454683&   13.12&    0.32& -0.79\\
2454237&   13.17&    0.28& -0.75&2454684&   13.04&    0.35& -0.65\\
2454246&   13.14&    0.33& -0.77&2454685&   13.14&    0.34& -0.71\\
2454259&   13.14&    0.32& -0.78&2454686&   13.19&    0.28& -0.68\\
2454260&   13.24&    0.28& -0.71&2454687&   13.20&    0.32& -0.67\\
2454261&   13.08&    0.33& -0.77&2454688&   13.16&    0.32& -0.78\\
2454263&   13.03&    0.36& -0.80&2454703&   13.16&    0.33& -0.86\\
2454265&   13.09&    0.32& -0.77&2454704&   13.12&    0.32& -0.77\\
2454266&   13.29&    0.22& -0.81&2454707&   13.11&    0.34& -0.82\\
2454267&   13.11&    0.34& -0.83&2454708&   13.16&    0.34& -0.86\\
2454269&   13.22&    0.28& -0.76&2454710&   13.20&    0.38& -0.76\\
2454270&   13.12&    0.31& -0.70&2454712&   13.23&    0.29& -0.61\\
2454290&   13.07&    0.35& -0.75&2454713&   13.10&    0.34& -0.77\\
2454291&   13.20&    0.26& -0.71&2454715&   13.12&    0.32& -0.79\\
2454292&   13.09&    0.36& -0.80&2454717&   13.18&    0.34& -0.77\\
2454293&   13.24&    0.30& -0.78&2454972&   13.15&    0.34& -0.77\\
2454296&   13.15&    0.35& -0.82&2454974&   13.19&    0.35& -0.77\\
2454298&   13.31&    0.31& -0.70&2454978&   13.18&    0.34& -0.72\\
2454299&   13.19&    0.27& -0.74&2454982&   13.17&    0.26& -0.75\\
2454300&   13.11&    0.28& -0.77&2455005&   13.13&    0.30& -0.82\\
2454301&   13.12&    0.31& -0.80&2455010&   13.15&    0.33& -0.82\\
2454302&   13.30&    0.25& -0.70&2455014&   13.16&    0.37& -0.77\\
2454303&   13.18&    0.34& -0.84&2455035&   13.16&    0.29& -0.68\\
2454304&   13.26&    0.31& -0.74&2455041&   13.05&    0.36& -0.82\\
2454316&   13.09&    0.35& -0.75&2455056&   13.17&    0.37& -0.80\\
2454326&   13.22&    0.25& -0.78&2455058&   13.11&    0.35& -0.74\\
2454328&   13.15&    0.28& -0.74&2455060&   13.25&    0.24& -0.74\\
2454330&   13.14&    0.30& -0.73&2455062&   13.18&    0.32& -0.73\\
2454331&   13.28&    0.23& -0.79&2455065&   13.14&    0.32& -0.74\\
2454332&   13.23&    0.26& -0.66&2455067&   13.15&    0.34& -0.78\\
2454335&   13.22&    0.28& -0.79&2455071&   13.16&    0.27& -0.75\\
2454361&   13.21&    0.32& -0.78&2455091&   13.03&    0.39& -0.78\\
2454363&   13.16&    0.34& -0.80&2455098&   13.09&    0.31& -0.73\\
2454375&   13.20&    0.35& -0.74&2455114&   13.28&    0.32& -0.68\\
2454378&   13.20&    0.30& -0.73&2455331&   13.17&    0.26& -0.68\\
2454391&   13.31&    0.28& -0.69&2455337&   13.21&    0.32& -0.78\\
2454566&   13.09&    0.40& -0.70&2455357&   13.18&    0.29& -0.72\\
2454586&   13.20&    0.26& -0.72&2455358&   13.32&    0.26& -0.72\\
2454587&   13.09&    0.40& -0.80&2455360&   13.14&    0.36& -0.74\\
2455366&   13.11&    0.34& -0.80&2455737&   13.10&    0.32& -0.78\\
2455389&   13.15&    0.30& -0.82&2455748&   13.15&    0.31& -0.75\\
2455393&   13.02&    0.35& -0.74&2455750&   13.19&    0.28& -0.73\\
2455412&   13.17&    0.30& -0.73&2455751&   13.13&    0.33& -0.76\\
2455416&   13.25&    0.30& -0.80&2455766&   13.11&    0.30& -0.74\\
2455418&   13.13&    0.33& -0.88&2455767&   13.14&    0.32& -0.73\\
2455421&   13.18&    0.32& -0.70&2455771&   13.01&    0.33& -0.75\\
2455422&   13.25&    0.28& -0.76&2455774&   13.09&    0.30& -0.73\\
2455423&   13.11&    0.35& -0.76&2455775&   13.09&    0.29& -0.70\\
2455445&   13.06&    0.31& -0.77&2455778&   13.10&    0.33& -0.76\\
2455448&   13.02&    0.33& -0.73&2455779&   13.04&    0.31& -0.73\\
2455454&   13.04&    0.30& -0.72&2455780&   13.03&    0.34& -0.78\\
2455712&   13.06&    0.36& -0.74&2455781&   13.17&    0.29& -0.78\\
2455717&   13.12&    0.33& -0.74&2455782&   13.17&    0.32& -0.84\\
2455719&   13.14&    0.31& -0.76&2455793&   13.15&    0.33& -0.75\\
2455720&   13.09&    0.30& -0.71&2455794&   13.10&    0.30& -0.74\\
2455721&   13.18&    0.32& -0.76&2455795&   13.13&    0.31& -0.75\\
2455736&   13.00&    0.33& -0.75&&&&\\
 \hline

\end{longtable}
\end{center}

}

\PZfig{10cm}{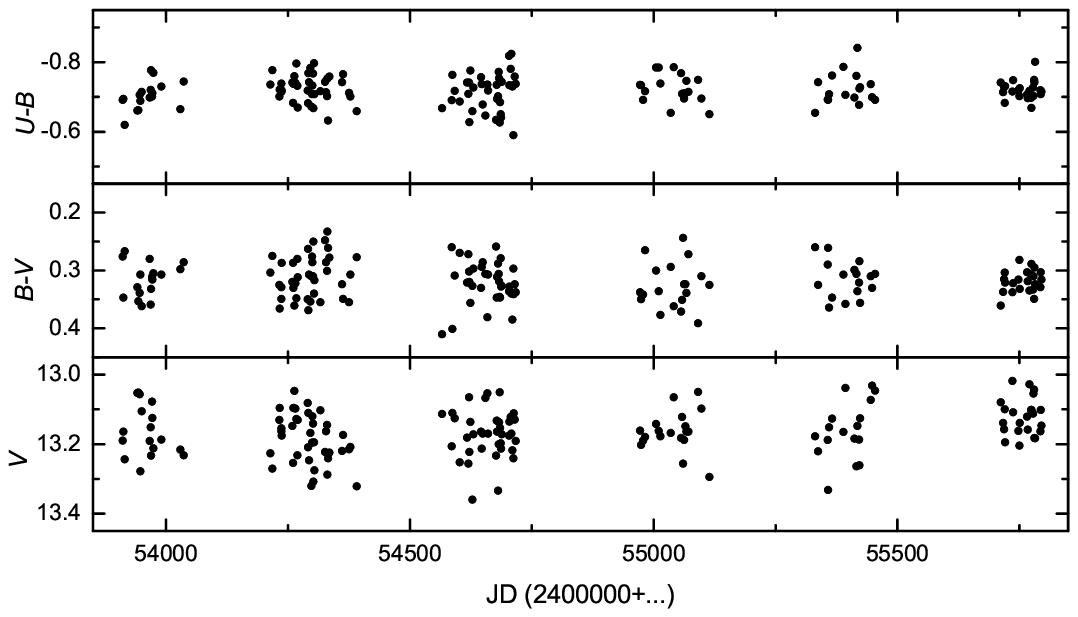}{Light and color curves of IRAS~19336--0400
in 2006-2011.}

\PZfig{10cm}{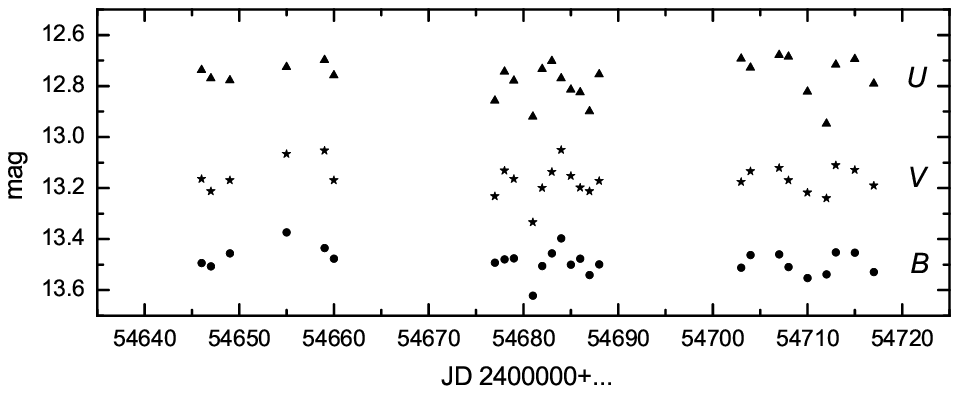}{\textit{U}, \textit{B}, \textit{V} curves
of
  IRAS~19336--0400 in the range JD 2454646-2454717.}

\PZfig{8cm}{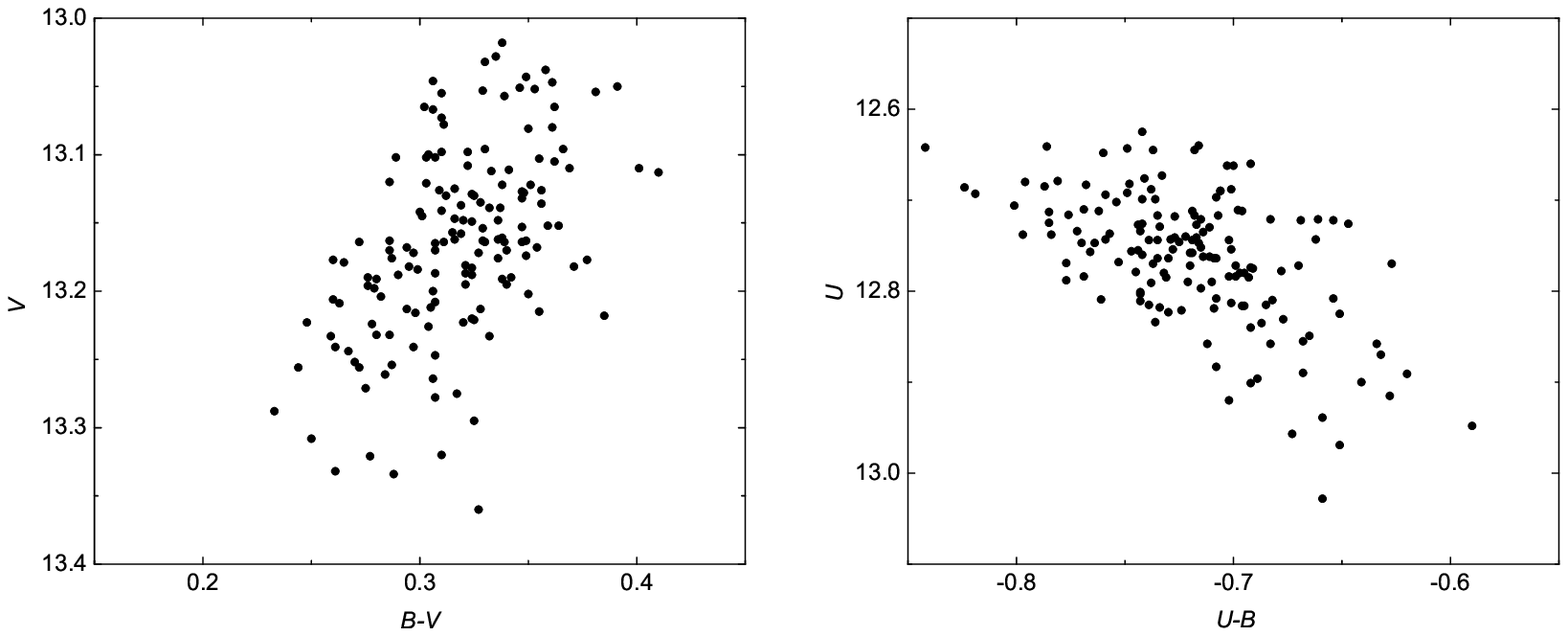}{Color--magnitude diagrams.}

\PZfig{9cm}{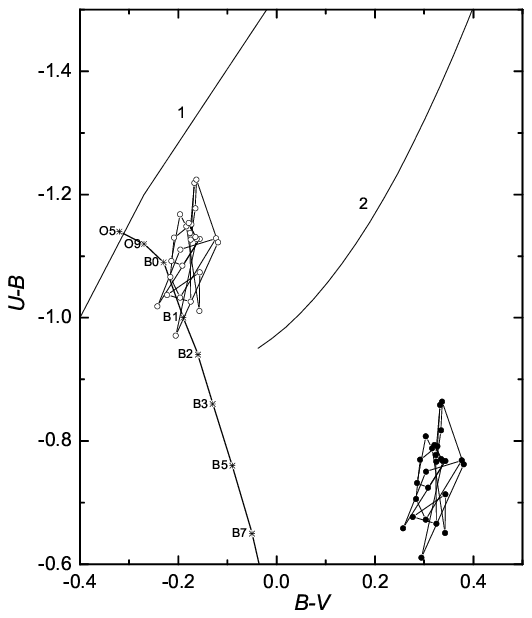}{Two--color diagram for the observations
presented in  Fig.~2. Colors of IRAS~19336--0400, observed and
corrected with $E_{B-V}=0.5$, are shown by points and open
circles, respectively. This diagram also displays the supergiant
sequence and theoretical $(U-B),(B-V)$ relations for emitting
plasma optically thin in Balmer continuum with $T_e=10\,000$~K,
$n_e=10^{10}$~cm$^{-3}$ (curve~1) and optically thick with
$T_e=15\,000$~K (curve~2) from Chalenko (1999).}

\PZfig{7cm}{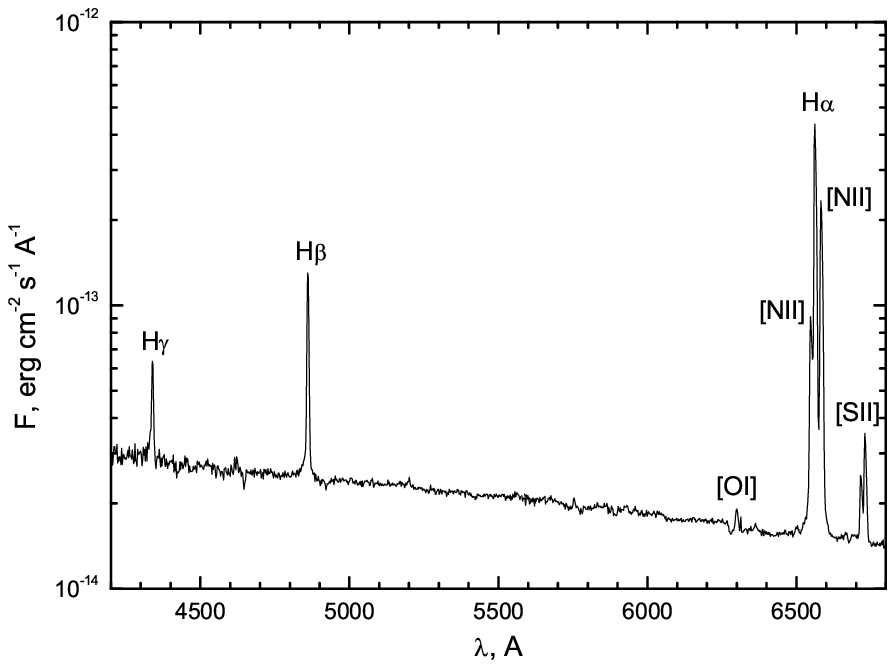}{IRAS~19336--0400 spectrum obtained on July
4, 2008.}

\PZfig{7cm}{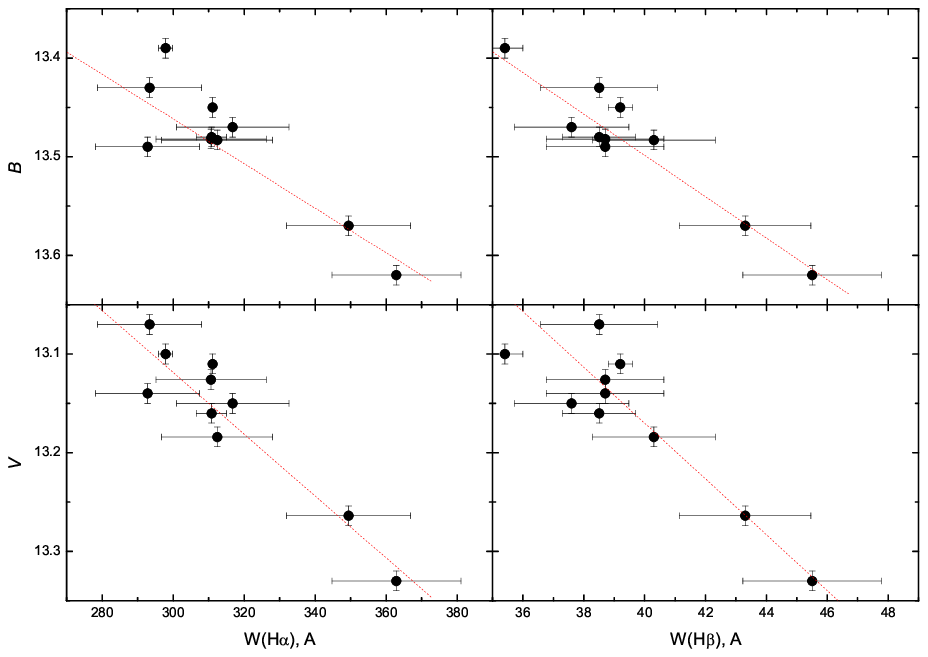}{Relations between H$\alpha$ and H$\beta$
emission line equivalent widths and brightness in \textit{B} and
    \textit{V}--bands.}

\newpage
\begin{table}
\begin{center}
\caption{Relative observed spectral line intensities scaled to
$F_{\textrm{H}\beta}=100$ and physical conditions for
IRAS~19336--0400.}\vspace{2mm}
\begin{tabular}{p{4cm}|p{2cm}|p{2.5cm}|p{3.5cm}} \hline \rule[-9pt]{0pt}{24pt} Parameters &
 This paper & Van de Steene et al. (1996)$^*$ & Pereira and Miranda (2007) \\
\hline \rule{0pt}{14pt}\hspace{-1.5mm}
$F_{\textrm{H}_\beta}\times10^{13}$, erg/cm$^{2}$/s
& 8.75 & -- & 1.8\\
4340 & 33 & 35.6 & 33\\
5755 & 1.4 & 1.8 & 2.5 \\
6548 & 92 & 98 & 82.7 \\
6563 & 510 & 516 & 578 \\
6584 & 261 & 250.5 & 300 \\
$R(S~II)$ & 0.5 & 0.5 & 0.5 \\
$c_{\textrm{H}_\beta}$ & 0.8 & 0.8 & -- \\
$E_{B-V}$ & 0.54 & 0.54 & $0.55^{**}$ \\
 \hline \rule{0pt}{14pt}\hspace{-1.5mm}
$T_e$, K & $7000^{+2500}_{-1000}$ & -- & $7600\pm600$ \\
$n_e$, cm$^{-3}$ & $\sim10^4$ & -- & $13000\pm3600$ \\

\hline
\end{tabular}
\end{center}
 $^*$ The paper contains line intensities corrected for
 interstellar reddening; we have recalculated them back to the
 observed ones using the extinction law of Seaton (1979) and
 $c_{\textrm{H}\beta}$ value obtained by the authors themselves.

 $^{**}$ Pereira and Miranda (2007) adduce $E_{B-V}$ value without
 mentioning what extinction law has been used; applying that of
 Seaton (1979) for $\textrm{H}_\alpha/\textrm{H}_\beta=5.78$ ratio
 gives $E_{B-V}=0.63$.

\end{table}

\end{document}